\documentclass[preprint,showpacs,showkeys,superscriptaddress,floatfix,
prd,amsmath,amssymb]{revtex4}

\usepackage[dvips]{color}                 %%% for colors in the graphs
\usepackage{graphicx}                     %%% for pictures and figures
\usepackage{amssymb}
\usepackage{amsmath}
\usepackage{xspace}                       %%% to get the correct spacing of
\usepackage{epsfig}
\begin{document}
%%%%%%%%%%%%%%%%%%%%%%%%%%%%%%%%%%%%

\title{Axial coupling constant of the nucleon for two flavors of
         dynamical quarks in finite and infinite volume}

\author{A. Ali Khan}
\affiliation{Institut f\"ur Theoretische Physik,
             Universit\"at Regensburg, 93040 Regensburg, Germany}
\author{M. G\"ockeler}
\affiliation{Institut f\"ur Theoretische Physik,
             Universit\"at Regensburg, 93040 Regensburg, Germany}
\author{Ph. H\"agler}
\affiliation{Physik-Department, Theoretische Physik,  
             Technische Universit\"at M\"unchen, 85747 Garching, Germany}
\author{T.R. Hemmert}
\affiliation{Physik-Department, Theoretische Physik,  
             Technische Universit\"at M\"unchen, 85747 Garching, Germany}
\author{R. Horsley}
\affiliation{School of Physics, University of Edinburgh,
             Edinburgh EH9 3JZ, UK}
\author{D. Pleiter}
\affiliation{John von Neumann-Institut f\"ur Computing NIC,
             Deutsches Elektronen-Synchrotron DESY, 15738 Zeuthen, Germany}
\author{P.E.L. Rakow}
\affiliation{Theoretical Physics Division,
             Department of Mathematical Sciences,
             University of Liverpool, Liverpool L69 3BX, UK}
\author{A. Sch\"afer}
\affiliation{Institut f\"ur Theoretische Physik,
             Universit\"at Regensburg, 93040 Regensburg, Germany}
\author{G. Schierholz}
\affiliation{John von Neumann-Institut f\"ur Computing NIC,
             Deutsches Elektronen-Synchrotron DESY, 15738 Zeuthen, Germany}
\affiliation{Deutsches Elektronen-Synchrotron DESY, 22603 Hamburg, Germany}
\author{T. Wollenweber}
\affiliation{Physik-Department, Theoretische Physik,  
             Technische Universit\"at M\"unchen, 85747 Garching, Germany}
\author{J.M. Zanotti}
\affiliation{School of Physics, University of Edinburgh,
             Edinburgh EH9 3JZ, UK}
\collaboration{QCDSF Collaboration} \noaffiliation
  
\begin{abstract}
We present data for the axial coupling constant $g_A$ of the nucleon
obtained in lattice QCD with two degenerate flavors of dynamical 
non-perturbatively improved Wilson quarks. The renormalization is also
performed non-perturbatively. For the analysis we give
a chiral extrapolation formula for $g_A$ based on the small scale 
expansion scheme of chiral effective field theory for two degenerate 
quark flavors. Applying this formalism in a finite volume we derive
a formula that allows us to extrapolate our data simultaneously to 
the infinite volume and to the chiral limit.
Using the additional lattice data in finite volume we are able to 
determine the axial coupling of the nucleon in the chiral limit 
without imposing the known value at the physical point. 
\end{abstract}



\maketitle

%%%%%%%%%%%%%%%%%%%%%%%%%%%%%%%%%%%
\section{Introduction}
%%%%%%%%%%%%%%%%%%%%%%%%%%%%%%%%%%%%%

The axial coupling constant $g_A$ of the nucleon
has been studied theoretically as well as experimentally for many years.
It can be defined as the value of the axial form factor of the nucleon
at vanishing 4-momentum transfer. More explicitly, one considers the
isovector axial current 
$A_\mu^{u-d} =  
\bar{u} \gamma_\mu \gamma_5 u - \bar{d} \gamma_\mu \gamma_5 d $,
where $u$ and $d$ denote the up and down quark fields, respectively.
We work in the limit of exact isospin invariance, i.e.\ $u$ and $d$
quarks are assumed to be degenerate in mass. The proton matrix 
element of this current has the form factor
decomposition
\begin{equation}
\langle p',s' | A_\mu^{u-d} | p,s \rangle
= \bar{u} (p',s') \left[ \gamma_\mu \gamma_5 G_A (q^2)
+ \gamma_5 \frac{q_\mu}{2 m_N} G_P (q^2) \right] u (p,s) \,.
\label{FF}
\end{equation}
Here $q = p'-p$ denotes the 4-momentum transfer and $u(p,s)$ is the proton 
spinor for momentum $p$ and spin vector $s$.
The states are normalized according to 
$\langle p',s' | p,s \rangle 
= (2 \pi)^3 \, 2p^0 \delta(\vec{p} - \vec{p}\, {}') \delta_{s s'}$, 
we take $s^2 = - m_N^2$, and $m_N$ is the nucleon mass. So $g_A = G_A(0)$
is determined by the forward matrix element
\begin{equation}
 \langle p,s | A_\mu^{u-d} | p,s \rangle = 2 g_A s_\mu \,.
\label{gadef}
\end{equation}

In parton model language, the forward matrix elements of the axial current
are related to the fraction of the nucleon spin carried by the quarks.
Denoting by $\Delta u$ ($\Delta d$) the contribution of the 
$u$ ($d$) quarks, one has 
\begin{equation}
 \langle p,s |\bar{u} \gamma_\mu \gamma_5 u  | p,s \rangle 
= 2 \Delta u \, s_\mu 
\end{equation}
and similarly for the $d$ quarks. Thus we can write 
$g_A = \Delta u - \Delta d$.

In this paper we report on new results for $g_A$ obtained by means of
Monte Carlo simulations of lattice QCD with two dynamical quark flavors.
While Eq.~(\ref{gadef}) lends itself immediately to an evaluation 
on the lattice, it is not yet possible to perform the simulations
at physical quark masses. Moreover, the size of the lattice is 
necessarily finite and in practice not very large. Thus, apart from the 
unavoidable continuum extrapolation, we have to cope with the thermodynamic
limit and the extrapolation to small (physical) quark masses, the so-called
chiral extrapolation. Guidance for these extrapolations is provided
by chiral effective field theory (ChEFT).

In its standard form, ChEFT describes low-energy QCD by means of an
effective field theory based on pion, nucleon, {\ldots} fields
taking into account the constraints imposed by (spontaneously
broken) chiral symmetry. As long as one stays in the $p$-regime of
ChEFT (with appropriate boundary conditions), the Lagrangian does not 
depend on the volume. So besides the quark-mass dependence 
the very same Lagrangian governs also the volume
dependence, and finite size effects can be calculated by evaluating the
theory in a finite (spatial) volume. Thus the finite volume does not
introduce any new parameters and the study of the finite size effects
yields an additional handle on the coupling constants of ChEFT.
 
There are several ways to treat baryons in ChEFT. Here we apply
the (non-relativistic) small scale expansion (SSE)~\cite{SSE}, 
which uses explicit pion, nucleon and $\Delta$(1232) degrees of freedom,
and extend the previous calculations of the quark-mass dependence of 
$g_A$~\cite{BFHM,HPW} in this scheme to finite volume. The dependence 
on the lattice spacing $a$ could be included~\cite{afin}
(for a review see Ref.~\cite{baer}), but we shall 
not consider this possibility here. 

A somewhat more phenomenological approach to the chiral extrapolation 
of $g_A$ (and other nucleon matrix elements) has been developed
in Ref.~\cite{DMT}. The volume dependence of such matrix elements
has also been studied by several methods~\cite{beane,DL,TALY} and we shall 
compare our procedure in some detail with that of Beane 
and Savage~\cite{beane}.

Preliminary results of our investigation have been presented in
Refs.~\cite{Lattice04,Lattice05}.

%%%%%%%%%%%%%%%%%%%%%%%%%%%%%%%%%%%%
\section{The simulations} \label{simu}
%%%%%%%%%%%%%%%%%%%%%%%%%%%%%%%%%%%%%%

The QCDSF and UKQCD collaborations have generated ensembles 
of gauge field configurations using $N_f=2$ non-perturbatively 
${\mathcal O}(a)$ improved Wilson quarks and Wilson's plaquette action 
for the gauge fields. The simulation parameters are listed in 
Table~\ref{tab:param} along with some auxiliary results needed later on. 
Note that we have two groups of three ensembles each which differ only 
in the volume (simulations 9, 10, 11 and 12, 13, 14). In this paper 
we shall not consider any partially quenched results, hence we set 
$\kappa = \kappa_{\mathrm {sea}}$.

\begin{table*}
\caption{Simulation parameters together with results for the force scale 
$r_0$ and the pion mass $m_\pi$ in lattice units. Also the value of the 
plaquette $u_0^4$ is given.}
\label{tab:param}
\begin{ruledtabular}
\begin{tabular}{rllllllll}
{} &  \multicolumn{1}{c}{Coll.} & \multicolumn{1}{c}{$\beta $}
& \multicolumn{1}{c}{$\kappa_{\mathrm {sea}}$}
& \multicolumn{1}{c}{volume} 
& \multicolumn{1}{c}{$r_0/a$}
& \multicolumn{1}{c}{$a m_\pi$}
& \multicolumn{1}{c}{$u_0^4$} \\
\hline
 1 &  QCDSF   & 5.20 & 0.1342  & $16^3 \times 32$ &
4.077(70)  &  0.5847(12) &  0.528994(58) \\
 2 &  UKQCD   & 5.20 & 0.1350  & $16^3 \times 32$ &
4.754(45)  &  0.4148(13) &  0.533670(40) \\
 3 &  UKQCD   & 5.20 & 0.1355  & $16^3 \times 32$ &
5.041(53)  &  0.2907(15) &  0.536250(30) \\[0.5cm]
 4 &  QCDSF   & 5.25 & 0.1346  & $16^3 \times 32$ &
4.737(50)  &  0.4932(10) &  0.538770(41) \\
 5 &  UKQCD   & 5.25 & 0.1352  & $16^3 \times 32$ &
5.138(55)  &  0.3821(13) &  0.541150(30) \\
 6 &  QCDSF   & 5.25 & 0.13575 & $24^3 \times 48$ &
5.532(40)  &  0.2556(5)  &  0.543135(15) \\[0.5cm]
 7 &  UKQCD   & 5.29 & 0.1340  & $16^3 \times 32$ &
4.813(82)  &  0.5767(11) &  0.542400(50) \\
 8 &  QCDSF   & 5.29 & 0.1350  & $16^3 \times 32$ &
5.227(75)  &  0.4206(9)  &  0.545520(29) \\
 9 &  QCDSF   & 5.29 & 0.1355  & $12^3 \times 32$ &
\multicolumn{1}{c}{-} & \multicolumn{1}{c}{-} & \multicolumn{1}{c}{-} \\
10 &  QCDSF   & 5.29 & 0.1355  & $16^3 \times 32$ &
\multicolumn{1}{c}{-} & \multicolumn{1}{c}{-} & \multicolumn{1}{c}{-} \\
11 &  QCDSF   & 5.29 & 0.1355  & $24^3 \times 48$ &
5.566(64)  &  0.3269(7)  &  0.547094(23) \\
12 &  QCDSF   & 5.29 & 0.1359  & $12^3 \times 32$ &
\multicolumn{1}{c}{-} & \multicolumn{1}{c}{-} & \multicolumn{1}{c}{-} \\
13 &  QCDSF   & 5.29 & 0.1359  & $16^3 \times 32$ &
\multicolumn{1}{c}{-} & \multicolumn{1}{c}{-} & \multicolumn{1}{c}{-} \\
14 &  QCDSF   & 5.29 & 0.1359  & $24^3 \times 48$ &
5.88(10)   &  0.2392(9)  &  0.548286(57) \\[0.5cm]
15 &  QCDSF   & 5.40 & 0.1350  & $24^3 \times 48$ &
6.092(67)  &  0.4030(4)  &  0.559000(19) \\
16 &  QCDSF   & 5.40 & 0.1356  & $24^3 \times 48$ &
6.381(53)  &  0.3123(7)  &  0.560246(10) \\
17 &  QCDSF   & 5.40 & 0.1361  & $24^3 \times 48$ &
6.714(64)  &  0.2208(7)  &  0.561281(8) \\
\end{tabular}
\end{ruledtabular}
\end{table*}

\begin{table*}
\caption{Bare and renormalized results for $g_A$. The values for $L$ 
and $m_\pi$ in physical units have been calculated using 
$r_0 = 0.467 \, \mbox{fm}$ together with $r_0/a$ at the respective 
quark masses.}
\label{tab:gabare}
\begin{ruledtabular}
\begin{tabular}{rlllllllllll}
{} & \multicolumn{1}{c}{$\beta $}
& \multicolumn{1}{c}{$\kappa_{\mathrm {sea}}$}
& \multicolumn{1}{c}{volume}
& \multicolumn{1}{c}{$L [\mbox{fm}]$}
& \multicolumn{1}{c}{$m_\pi [\mbox{GeV}]$}
& \multicolumn{1}{c}{$g_A^{\mathrm {bare}}$} 
& \multicolumn{1}{c}{$g_A$} \\
\hline
 1 & 5.20 & 0.1342  & $16^3 \times 32$ & 1.84 & 1.007(17)
   & 1.452(17) & 1.185(16) \\
 2 & 5.20 & 0.1350  & $16^3 \times 32$ & 1.57 & 0.8332(83)
   & 1.514(29) & 1.201(24) \\
 3 & 5.20 & 0.1355  & $16^3 \times 32$ & 1.48 & 0.6192(73)
   & 1.396(36) & 1.088(29) \\[0.5cm]
 4 & 5.25 & 0.1346  & $16^3 \times 32$ & 1.58 & 0.987(11)
   & 1.442(13) & 1.176(12) \\
 5 & 5.25 & 0.1352  & $16^3 \times 32$ & 1.45 & 0.8295(93)
   & 1.438(20) & 1.148(17) \\
 6 & 5.25 & 0.13575 & $24^3 \times 48$ & 2.03 & 0.5975(45)
   & 1.456(10) & 1.1398(98) \\[0.5cm]
 7 & 5.29 & 0.1340  & $16^3 \times 32$ & 1.55 & 1.173(20)
   & 1.437(12) & 1.207(12) \\
 8 & 5.29 & 0.1350  & $16^3 \times 32$ & 1.43 & 0.929(13)
   & 1.409(12) & 1.143(11) \\
 9 & 5.29 & 0.1355  & $12^3 \times 32$ & 1.01 & \multicolumn{1}{c}{-}
   & 1.181(60) & 0.942(48) \\
10 & 5.29 & 0.1355  & $16^3 \times 32$ & 1.34 & \multicolumn{1}{c}{-}
   & 1.364(25) & 1.087(21) \\
11 & 5.29 & 0.1355  & $24^3 \times 48$ & 2.01 & 0.7688(90)
   & 1.459(11) & 1.163(11) \\
12 & 5.29 & 0.1359  & $12^3 \times 32$ & 0.95 & \multicolumn{1}{c}{-}
   & 0.97(10)  & 0.763(79) \\
13 & 5.29 & 0.1359  & $16^3 \times 32$ & 1.27 & \multicolumn{1}{c}{-}
   & 1.253(45) & 0.985(36) \\
14 & 5.29 & 0.1359  & $24^3 \times 48$ & 1.91 & 0.594(10)
   & 1.413(22) & 1.111(18) \\[0.5cm]
15 & 5.40 & 0.1350  & $24^3 \times 48$ & 1.84 & 1.037(11)
   & 1.4737(74) & 1.2234(88) \\
16 & 5.40 & 0.1356  & $24^3 \times 48$ & 1.76 & 0.8420(72)
   & 1.451(11) & 1.180(11) \\
17 & 5.40 & 0.1361  & $24^3 \times 48$ & 1.67 & 0.6264(63)
   & 1.410(20) & 1.127(17) \\
\end{tabular}
\end{ruledtabular}
\end{table*}

When we want to compare (fit) our results with formulae from ChEFT we 
need all numbers in physical units, i.e.\ we have to fix the scale.
This is usually done with the help of the force scale $r_0$~\cite{r0}
derived from the heavy-quark potential. From phenomenology we know
that $r_0 \approx 0.5 \, \mbox{fm}$ in the real world. Unfortunately,
it is not easy to determine this value precisely because it is related
to measurable quantities only through the use of potential models.

Yet this is not the only problem with setting the scale.
In the analysis of quenched simulations it is common practice to 
identify $r_0$ as extracted from the Monte Carlo data with the 
physical $r_0$. However, the latter refers to a world where the sea 
quarks have their physical masses while in the quenched simulations 
they are infinitely heavy. Still, for many quantities quenched 
calculations lead to results that agree surprisingly well with experiment.

In simulations with dynamical quarks additional difficulties arise.
In this case $r_0/a$ depends not only on $\beta$ as in the quenched
approximation, but also on $\kappa$. Then one can either 
identify the value for $r_0$ obtained at the given 
($\beta$, $\kappa$) combination with the physical $r_0$ or one can 
attempt an extrapolation of $r_0/a$ to the chiral limit, e.g.\ at 
fixed $\beta$, and use the extrapolated value to set the scale. 
The first procedure is analogous to the standard approach in the quenched 
case and ensures a continuous connection with the quenched results
as the quark mass tends to infinity. The second procedure, on the other
hand, has the virtue of leading to a mass-independent lattice spacing $a$.
It is however to be noted that, strictly speaking, the chiral 
extrapolation should be performed at fixed 
$\tilde{g}_0^2 = (1 + b_g am) g_0^2$ (see Ref.~\cite{luescher}) 
and not at fixed $\beta$, although the difference seems to be rather 
small in practice.

We have determined $r_0$ in physical units from the nucleon mass. 
Of course, this can only be done after a chiral extrapolation
to the physical pion mass. Encouraged 
by our comparison of nucleon mass data in different volumes with chiral
perturbation theory~\cite{finitemass}, we have used the extrapolation
procedure described in~\cite{finitemass} as Fit~1 to recent nucleon masses 
obtained by the CP-PACS and JLQCD collaborations along with updated
masses from the QCDSF-UKQCD collaboration. 
Varying the assumed physical value of $r_0$ one can make the 
fit curve pass through the physical point, which 
happens for $r_0=0.467 \, \mbox{fm}$. 
This number is consistent with a recent lattice
calculation of $f_\pi$~\cite{fpi}, giving $r_0=0.475(25) \,
\mbox{fm}$. A similar result for $r_0$ was also quoted in~\cite{aubin}
taking as input level splittings in the $\Upsilon$ spectrum. 
In the following we shall use $r_0=0.467 \, \mbox{fm}$, but
for comparison we shall also consider $r_0 = 0.5 \, \mbox{fm}$.

We compute $g_A$ from forward proton matrix elements (\ref{gadef})
of the flavor-nonsinglet axial vector current at $\vec{p} = \vec{0}$.
The required bare matrix elements are extracted from ratios of 3-point 
functions over 2-point functions in the standard fashion. Our results
for $g_A^{\mathrm {bare}}$ are collected in Table~\ref{tab:gabare}.
Compared to the computation of hadron masses, additional difficulties may 
arise in the calculation of nucleon matrix elements such as $g_A$: 
In general there are quark-line disconnected contributions, which are 
hard to evaluate, the operators must be improved and renormalized etc. 
Fortunately, in the limit of exact isospin invariance, which is taken 
in our simulations, all disconnected contributions cancel in $g_A$, 
because it is a flavor-nonsinglet quantity. So we do not have to 
worry about any disconnected contributions.

The axial vector current is $\mathcal O(a)$ improved by adding
the derivative of the pseudoscalar density with the appropriate
coefficient:
\begin{equation}
A_\mu^{\mathrm {imp}}(x) =  \bar{q}(x) \gamma_\mu \gamma_5 q(x)
+  a c_A^{\mathrm {imp}}  \partial_\mu  \bar{q}(x) \gamma_5 q(x) \,.
\end{equation}
Hence the improvement term does not contribute in forward matrix elements
such as (\ref{gadef}). We have denoted the improvement 
coefficient (usually called $c_A$) by $c_A^{\mathrm {imp}}$ in order to
avoid confusion with a coupling constant appearing later.

The renormalized improved axial vector current can be written as
\begin{equation}
A_\mu = Z_A \left( 1 + b_A a m \right) A_\mu^{\mathrm {imp}}
\end{equation}
with the bare quark mass 
$m = \left( 1/\kappa - 1/\kappa_c \right)/(2a) $. The values of
$\kappa_c$ used here can be found in Table~\ref{tab:za}.
The coefficient $b_A$ is required to maintain $O(a)$ improvement also
for non-vanishing quark mass. We are not aware of a non-perturbative
evaluation of this coefficient for our action. Hence we have to resort
to perturbation theory. A one-loop calculation yields~\cite{sint}
\begin{equation}
b_A = 1 + b_A^{(1)} g_0^2 + \mathcal O(g_0^4) \: ,\: 
b_A^{(1)} = 0.11414(4) \, C_F \,,
\end{equation}
where $C_F = 4/3$ in QCD and $g_0$ denotes the bare coupling with
$\beta = 6/g_0^2$. We shall
use the tadpole improved version of this result, i.e.\ we take~\cite{lanl}
\begin{equation}
b_A^{\mathrm {TI}} = 
u_0^{-1} \left( 1 + \left(b_A^{(1)} - \frac{1}{12} \right)  
                                       \frac{g_0^2}{u_0^4} \right) \,.
\end{equation}
Here $u_0$ is the fourth root of the expectation value of the plaquette
with the perturbative expansion 
\begin{equation}
u_0 = 1 - \frac{1}{12} g_0^2 + \mathcal O(g_0^4) \,.
\end{equation}

\begin{figure}[htb]
\begin{center}
\epsfig{file=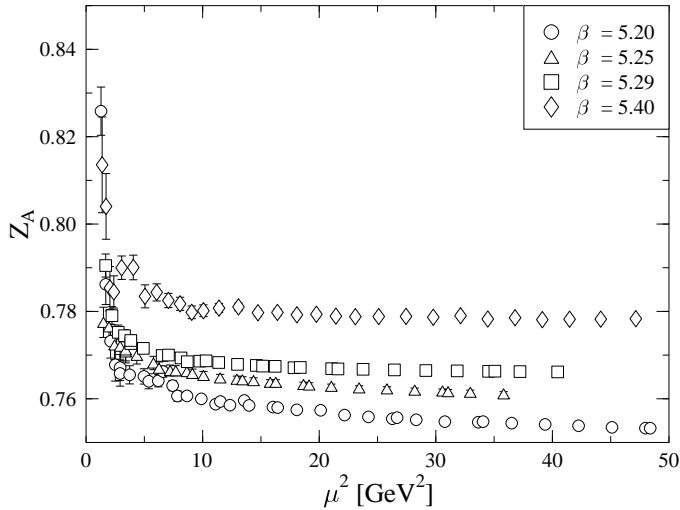,width=10cm}
\end{center}
\vspace*{-1.0cm}
\caption{The renormalization constant $Z_A$ plotted versus the square of the
renormalization scale $\mu$.}
\label{fig:za}
\end{figure}

While the coefficient $b_A$ will be computed in tadpole improved one-loop
perturbation theory, we calculate the renormalization factor $Z_A$ 
non-perturbatively by means of the Rome-Southampton method~\cite{rimom,reno}. 
Thus $Z_A$ is first obtained in the so-called RI$^\prime$-MOM 
scheme. Using continuum perturbation~\cite{gracey} theory we 
switch to the $\overline{\mbox{MS}}$ scheme. For sufficiently large
renormalization scales $\mu$, $Z_A$ should then be independent of $\mu$.
However, unless $\mu \ll 1/a$ lattice artefacts may spoil this behavior.
Since our scales do not always satisfy this criterion, we try to correct 
for this mismatch by subtracting the lattice artefacts perturbatively with
the help of boosted one-loop lattice perturbation theory. 
As Fig.~\ref{fig:za} shows, some lattice artefacts still remain, but
we can nevertheless estimate $Z_A$. As our central value we take the result
obtained at $\mu^2 = 5 \, \mbox{GeV}^2$ while the difference to the result
at $\mu^2 = 10 \, \mbox{GeV}^2$ gives us an estimate of the (systematic) error.
Adding the statistical error in quadrature we arrive at the numbers given
in Table~\ref{tab:za}, where we also compare our 
results with a recent determination of $Z_A$ by the ALPHA 
collaboration~\cite{alpha}. While the numbers differ significantly at
the lower $\beta$ values there is a tendency towards agreement as $\beta$
grows. This suggests that the differences are mainly due to lattice 
artefacts and will not influence the continuum limit of renormalized
quantities.

\begin{table*}
\caption{Values of $Z_A$ from this work and from the ALPHA collaboration.
The latter numbers were obtained from Eq.~(3.6) in Ref.~\cite{alpha}, where
an error decreasing from 0.01 at $\beta = 5.2$ to 0.005 at $\beta = 5.7$
is ascribed to them. In addition we give $\kappa_c$ and the values of 
$r_0/a$ extrapolated to the chiral limit (from~\cite{lambda}). }
\label{tab:za}
\begin{ruledtabular}
\begin{tabular}{lllll}
$\beta$           &  5.20     & 5.25     & 5.29     & 5.40      \\
\hline
$Z_A$ (this work) &  0.765(5) & 0.769(4) & 0.772(4) & 0.783(4)  \\
$Z_A$ (ALPHA)     &  0.719    & 0.734    & 0.745    & 0.767  \\
$\kappa_c$        &  0.136008(15) & 0.136250(7) & 0.136410(9) & 0.136690(22) \\
$(r_0/a)_{\mathrm {ch.l.}}$ & 5.455(96) & 5.885(79) & 6.254(99) & 7.39(26) \\
\end{tabular}
\end{ruledtabular}
\end{table*}

\begin{figure}
\begin{center}
\epsfig{file=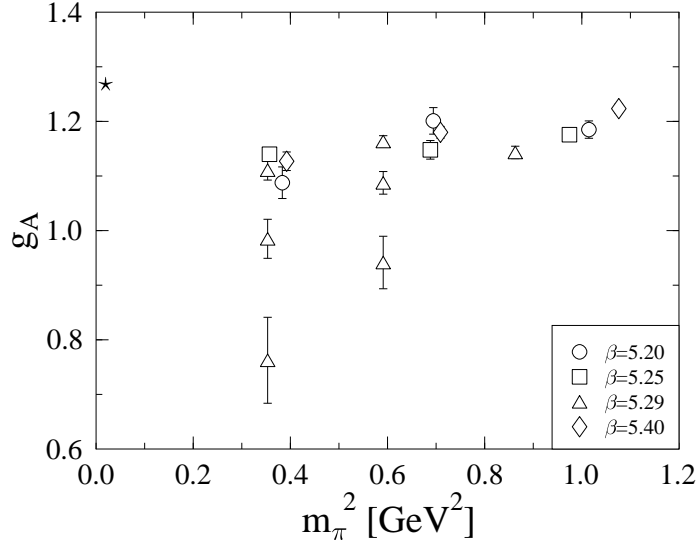,width=10cm}
\end{center}
\caption{Results for $g_A$. For
 $(\beta,\kappa) = (5.29,0.1355)$ and $(5.29,0.1359)$
we show the results obtained on three different spatial volumes. The smallest
box size $L$, leading to the smallest $g_A$, is about $1.0 \, \mbox{fm}$,
while the middle size is approximately $1.3 \, \mbox{fm}$. All other
volumes are larger. The star represents the physical point.}
\label{fig:gadat}
\end{figure}

Our results for $g_A$ renormalized in the way just described are given in
the last column of Table~\ref{tab:gabare} and are plotted in 
Fig.~\ref{fig:gadat}. Note that the error on $g_A$ is dominated by
the statistical error of the bare values while the uncertainty of $Z_A$
contributes only comparatively little. We have taken $m_\pi$ from the 
largest available lattice at each ($\beta$, $\kappa$)
combination. The scale has been set by means of the force parameter $r_0$
with $r_0 = 0.467 \, \mbox{fm}$, and $r_0/a$ has been taken at the given
quark mass. Obviously there are considerable finite size effects. 
A qualitatively similar volume dependence has already been 
observed in quenched simulations~\cite{sasaki}. Note that the 
``large volume'' results for $g_A$ obtained at our smallest quark masses 
($m_\pi \approx 600 \, \mbox{MeV}$) for all four $\beta$ values
lie very close together indicating that discretization effects are small.

Our results may be compared with other evaluations of $g_A$ in
dynamical simulations~\cite{dolgov,ohta,MIT}. They all differ in the 
lattice actions employed: In Ref.~\cite{dolgov} unimproved Wilson 
fermions are used, the RBCK collaboration~\cite{ohta} works with 
domain wall fermions, and the most recent LHPC investigation~\cite{MIT} 
chooses a hybrid approach with domain wall valence quarks on improved
staggered sea quarks. All these studies obtain a rather weak quark-mass
dependence of $g_A$ in agreement with our findings. However, most of
their values lie somewhat above ours. At the moment, the reason for these
discrepancies is not yet clear. One possibility is the different treatment
of the renormalization. The perturbative renormalization employed in
Ref.~\cite{dolgov} entails a relatively large uncertainty because
it relies on a one-loop calculation only and in general lattice perturbation 
theory does not converge very rapidly. References~\cite{ohta,MIT} 
exploit the fact that in the framework of domain wall fermions 
there are five-dimensional (partially) conserved axial and vector 
currents. One should however keep in mind that they are exactly 
conserved only in the limit of an infinite
fifth dimension, while the actual simulations are necessarily performed
with this extension kept finite.

Note that using the $Z_A$ results of the ALPHA
collaboration we would have obtained even smaller numbers. As indicated 
above, this difference may be expected to disappear in the continuum limit.
However, from the comparison of our results at different values of $\beta$
we have the impression that lattice artefacts are rather small in the data
renormalized with our $Z_A$ values (see Fig.~\ref{fig:gadat}). 
This is less so when the $Z_A$ of the ALPHA
collaboration is used. In any case, there is still some work to be 
done before we can be sure that all systematic uncertainties are 
under control.
 
%%%%%%%%%%%%%%%%%%%%%%%%%%%%%%%%%%%%%%%%%
\section{Chiral effective field theory (infinite volume)}
%%%%%%%%%%%%%%%%%%%%%%%%%%%%%%%%%%%%%%%%%%
\label{sec:cheft}

\begin{figure}[ht]
  \begin{center}
\epsfig{file=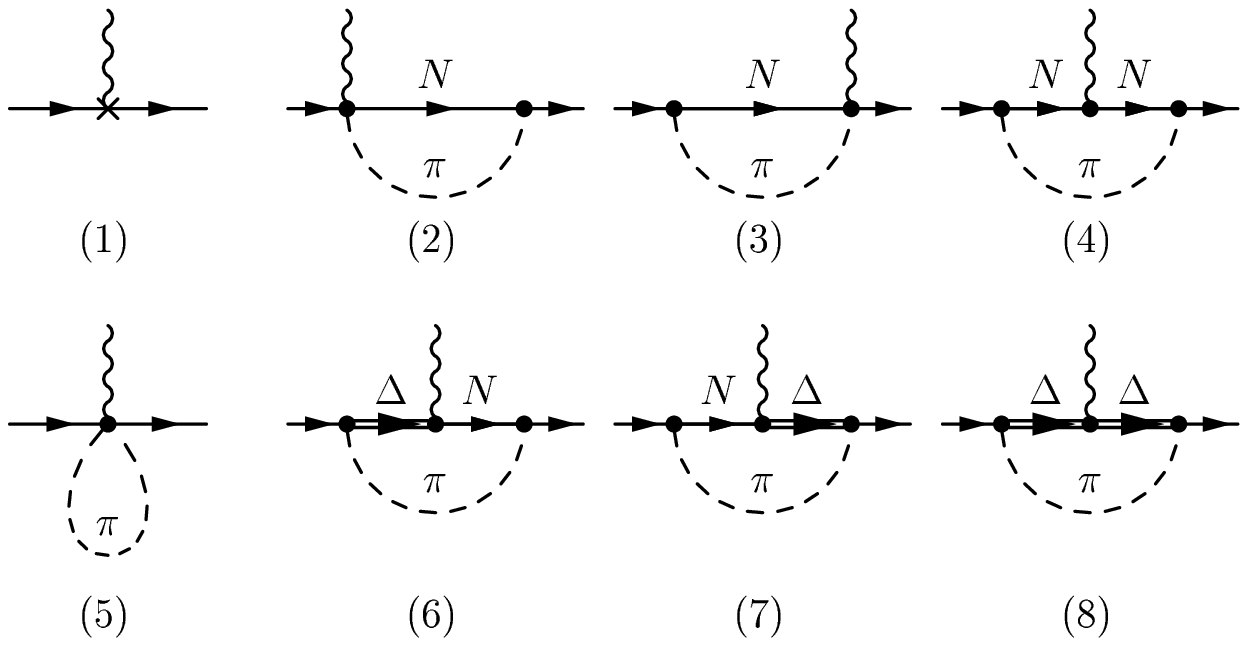,width=0.7\textwidth}
    \caption{Diagrams contributing to the quark-mass dependence of 
    $g_A$ up to ${\cal O}(\epsilon^3)$ in the SSE scheme.}
  \label{diagrams}
  \end{center}
\end{figure}

At low energies one can evaluate the matrix element~(\ref{FF}) 
of the isovector axial current of the nucleon within a low energy 
effective theory of QCD utilizing the methods of ChEFT. In the forward limit 
(Eq.(\ref{gadef})) one then obtains the axial coupling constant 
$g_A$. In the SSE formalism the results appear as expansions in powers
of a small parameter $\epsilon$, which collectively denotes small pion
four-momenta, the pion mass, baryon three-momenta and the nucleon-$\Delta$
mass splitting in the chiral limit~\cite{SSE}.
At ${\mathcal O}(\epsilon^3)$ one has to 
evaluate 8 diagrams \cite{BFHM}, which are displayed in Fig.~\ref{diagrams},  
involving nucleon, pion and $\Delta$(1232) degrees of freedom. When 
ChEFT calculations of nucleon properties using just pions and nucleons 
as the active degrees of freedom are extended to ChEFT 
calculations of the same quantities also employing explicit
$\Delta$(1232) degrees of freedom, one obtains renormalizations of the 
chiral limit couplings of the nucleon by polynomial terms
$\sim \left(\Delta_0\right)^n$, where $\Delta_0$ denotes the (finite) 
N-$\Delta$ mass splitting in the chiral limit. This would naively 
imply that the chiral limit properties of a nucleon differ between 
ChEFT schemes with and without resonance degrees of freedom. In the SSE 
formalism such an unphysical scenario is automatically avoided  because 
$\Delta_0$ (by construction) is treated  as a quantity of 
${\cal O}(\epsilon)$ leading to extra terms in the nucleon Lagrangian 
which cancel the contributions causing this behavior. We refer to 
Ref.~\cite{HW} for a detailed example of such a decoupling construction. 

Implementing this decoupling of the $\Delta$ resonance near the chiral 
limit leads to the result~\cite{HPW}:
\begin{eqnarray}
g_A^{SSE}(\infty)&=&g^0_A-\frac{(g^0_A)^3m_\pi^2}{16\pi^2F_\pi^2}
                     +4\left\{C^{SSE}(\lambda)
                     +\frac{c_A^2}{4\pi^2 F_\pi^2}\left[\frac{155}{972}\, g_1
                     -\frac{17}{36}\, g^0_A\right]+\gamma^{SSE}
                     \ln{\frac{m_\pi}{\lambda }}\right\}m_\pi^2\nonumber\\
                  & &{}+\frac{4c_A^2
                     g^0_A}{27 \pi F_\pi^2 \Delta_0}m_\pi^3
                     +\frac{8}{27\pi^2 F_\pi^2}\;c_A^2 g^0_A m_\pi^2
                     \sqrt{1-\frac{m_\pi^2}{\Delta_0^2}}\,\ln{R}\nonumber\\
                  & &{}+\frac{c_A^2\Delta_0^2}{81\pi^2 F_\pi^2}\left(25g_1-
                     57g_A^0\right)\left\{\ln\left[\frac{2\Delta_0}{m_\pi}
                     \right]-\sqrt{1-\frac{m_\pi^2}{\Delta_0^2}}\ln R\right\}
                     +{\mathcal O}(\epsilon^4)
\label{gasse}
\end{eqnarray}
with
\begin{eqnarray}
\gamma^{SSE}&=&\frac{1}{16\pi^2 F_\pi^2}\left[\frac{50}{81}\,c_A^2 g_1
-\frac{1}{2}\,g_A^0-\frac{2}{9}\,c_A^2g_A^0-(g_A^0)^3\right]\;,
                                   \nonumber\\
  R&=&\frac{\Delta_0}{m_\pi }+\sqrt{\frac{\Delta_0^2}{m_\pi^2}-1} \,.
\label{R}
\end{eqnarray}
In the SSE counterterm combination~\cite{HPW}
\begin{equation}
C^{SSE}(\lambda) = B_9^r(\lambda)^{SSE}-2g_A^0B_{20}^r(\lambda)^{SSE}  
\label{cSSE}
\end{equation}
$\lambda$ denotes the scale utilized in dimensional regularization, 
which can be freely chosen as the ChEFT results of course do not 
depend on the employed regularization scheme or scale. In order to make this 
scale independence explicit, the $\lambda$ dependence of all the couplings 
(controlled by their $\beta$ functions) has of course also to be taken 
into account (see Table~1 in Ref.~\cite{BFHM} for details). 

In Eqs.~(\ref{gasse}) and (\ref{R}) $g_A^0$ denotes the axial 
coupling constant of the nucleon in the chiral limit. The parameters 
$F_\pi$ and $\Delta_0$ are the pion decay constant (with the physical 
value $92.4 \, \mbox{MeV}$) and the real 
part of the $N\Delta$ mass splitting, whereas $c_A$ and $g_1$ denote the 
leading axial $N\Delta$ and $\Delta \Delta$ couplings, 
respectively. It is understood that these four parameters are also 
taken in the chiral limit.

Throughout this work we have converted the 
quark-mass dependence of $g_A$ into a dependence on the mass of the 
lowest lying $0^-$ boson (i.e.\ the ``pion'') in the theory via the relation
\begin{equation}
m_\pi^2 = 2\,B_0\,\hat{m}\left[1+{\mathcal O}(\hat{m})\right] \,, 
\label{GOR}
\end{equation}
where $\hat{m}=\left(m_u+m_d\right)/2$ denotes the average of the $u$ 
and $d$ quark mass and the parameter 
$B_0=-\langle 0|\bar{q}q|0\rangle/F_\pi^2$  measures the size of the 
SU(2) chiral condensate in the chiral limit. The leading term in 
Eq.~(\ref{GOR}) corresponds to the GOR-relation \cite{GOR}. The 
indicated higher order quark-mass terms only start contributing beyond the 
${\mathcal O}(\epsilon^3)$ considered here. We also note that Eq.~(\ref{GOR}) 
is consistent with the available lattice QCD data for the quark-mass
dependence of the pion mass (see, e.g., Fig.~1 in Ref.~\cite{pion}). 

Here we employ Eq.~(\ref{gasse}) in a different representation:
 \begin{eqnarray}
g_A\left(\infty\right)&=&g_A^0+\left[4\,B_9^r(\lambda)
-8\,g_A^0B_{20}^r(\lambda) -\frac{(g_A^0)^3}{16\pi^2F_\pi^2}
    -\frac{25c_A^2g_1}{324\pi^2F_\pi^2} 
  +\frac{19c_A^2g_A^0}{108\pi^2F_\pi^2}\right]m_\pi^2 \nonumber \\
    & & {} -\frac{m_\pi^2}{4\pi^2 F_\pi^2}\left[(g^0_A)^3
            +\frac{1}{2}\,g^0_A\right]\ln{\frac{m_\pi}{\lambda }} 
   +\frac{4c_A^2g_A^0}{27\pi\Delta_0F_\pi^2}\,m_\pi^3 \nonumber \\
   & &{} +\left[25c_A^2g_1\Delta_0^2-57c_A^2g_A^0\Delta_0^2
                        -24c_A^2g_A^0m_\pi^2\right]
      \frac{\sqrt{m_\pi^2-\Delta_0^2}}{81\pi^2F_\pi^2\Delta_0}
                  \arccos\frac{\Delta_0}{m_\pi}\nonumber \\
    & &{} +\frac{25c_A^2g_1\left(2\Delta_0^2-m_\pi^2\right)}{162\pi^2F_\pi^2}
                                              \ln\frac{2\Delta_0}{m_\pi}
    +\frac{c_A^2g_A^0\left(3m_\pi^2-38\Delta_0^2\right)}{54\pi^2F_\pi^2}
                                             \ln\frac{2\Delta_0}{m_\pi}
               +{\mathcal O}(\epsilon^4)\,. \label{gaSSEneu}
\end{eqnarray}
In Eq.~(\ref{gaSSEneu}) we have analytically continued the logarithms of 
Eq.~(\ref{gasse}) to the region $m_\pi>\Delta_0$, because that is the region
where our Monte Carlo data lie. Furthermore we have introduced the 
couplings $B_9^r(\lambda)$ and $B_{20}^r(\lambda)$ used in heavy baryon
chiral perturbation theory (HBChPT), which can be calculated from the 
SSE couplings via the relations
 \begin{eqnarray}
B_9^r(\lambda)^{SSE}     &=&B_9^r(\lambda)
  -\frac{115 c_A^2 g_1}{1944\pi^2F_\pi^2}
  -\frac{25 c_A^2 g_1}{648\pi^2F_\pi^2} \ln \frac{2\Delta_0}{\lambda} \,, 
  \label{b9} \\
B_{20}^r(\lambda)^{SSE}&=&B_{20}^r(\lambda)
 -\frac{35 c_A^2}{432\pi^2F_\pi^2}
 -\frac{c_A^2}{144\pi^2F_\pi^2} \ln \frac{2\Delta_0}{\lambda} 
  \label{b20} \,.
\end{eqnarray}
Thus we are able to exploit directly the empirical information
on these couplings (see Eq.~(\ref{bphen}) below) obtained from a HBChPT
calculation of $\pi N \to \pi \pi N$ in Ref.~\cite{Nadia}.
Another advantage of utilizing the  couplings $B_9^r(\lambda)$ and 
$B_{20}^r(\lambda)$ becomes clear when we study the chiral limit 
behavior of Eq.~(\ref{gaSSEneu}):
\begin{eqnarray}
g_A(\infty)&=&g^0_A-\frac{(g^0_A)^3}{16\pi^2F_\pi^2}\,m_\pi^2
+4\left[B_9^r(\lambda)-2\,g_A^0B_{20}^r(\lambda)\right]m_\pi^2
 -\frac{m_\pi^2}{4\pi^2 F_\pi^2}
    \left[(g^0_A)^3+\frac{1}{2}\,g^0_A\right]\ln{\frac{m_\pi}{\lambda }} 
\nonumber \\
                            & &{}+{\mathcal O}(m_\pi^3)\,.
\label{gAchi}
\end{eqnarray}
In Eq.~(\ref{gAchi}) it becomes manifest that the couplings 
defined in Eqs.~(\ref{b9}) and (\ref{b20}) ensure that the 
${\mathcal O}(\epsilon^3)$ SSE 
result of Eq.~(\ref{gaSSEneu}) displays the same chiral limit 
behavior as the ${\mathcal O}(p^3)$ HBChPT 
result for the quark-mass dependence of $g_A$ given in Ref.~\cite{BKM}. 
In Ref.~\cite{HPW} it was made sure that $g_A^0$ and 
the leading non-analytic quark-mass dependence $\sim \ln m_\pi$ agree 
in the HBChPT and the SSE calculation. Here this mapping of the
two field theories is extended to the terms $\sim m_\pi^2$. 
A discussion of this point will be given in Ref.~\cite{future}.

However, we want to emphasize again that the ${\mathcal O}(\epsilon^3)$ SSE
results of Refs.~\cite{BFHM,HPW} (i.e.\ Eq.~(\ref{gasse})) 
and our Eq.~(\ref{gaSSEneu}) are all equivalent and provide identical results, 
they only differ in the definitions of the employed counterterms.

%%%%%%%%%%%%%%%%%%%%%%%%%%%%%%%%%%%%%%%%%%%%%%
\section{Chiral effective field theory (finite volume)}
%%%%%%%%%%%%%%%%%%%%%%%%%%%%%%%%%%%%

Applying the methods discussed in Ref.~\cite{finitemass} 
(for a recent more detailed investigation of these finite volume
corrections see Ref.~\cite{musch}) 
we can now repeat the calculation of $g_A$ in the SSE scheme for a finite
spatial cubic box of length $L$ in order to obtain the volume 
dependence of $g_A$. We define
\begin{equation}
g_A\left(L\right) = g_A\left(\infty\right)+\Delta g_A\left(L\right) \,. 
\label{gafinite}
\end{equation}
For $g_A\left(\infty\right)$ we utilize the ${\mathcal O}(\epsilon^3)$ SSE 
result of Eq.~(\ref{gaSSEneu}). From the diagrams of Fig.~\ref{diagrams} 
we obtain for $\Delta g_A\left(L\right)$ to ${\mathcal O}(\epsilon^3)$ 
in SSE 
\begin{eqnarray}
\Delta g_A\left(L\right) &=& -\,\frac{g_A^0 m_\pi^2}{4\pi^2F_\pi^2}
{\sum_{\vec{n}}}' \frac{K_1\left(L |\vec{n}| m_\pi\right)}{L |\vec{n}| m_\pi} 
\nonumber  \\
 &  & {}+\frac{\left(g_A^0\right)^3 m_\pi^2}{6\pi^2F_\pi^2} 
{\sum_{\vec{n}}}'  \left[K_0\left(L |\vec{n}| m_\pi\right)
     -\frac{K_1\left(L |\vec{n}| m_\pi\right)}{L |\vec{n}| m_\pi} \right] 
\nonumber \\ 
 &   &{}+\frac{25c_A^2g_1}{81\pi^2F_\pi^2}\int_0^\infty dy\, y
{\sum_{\vec{n}}}'  \left[K_0\left(L|\vec{n}| f(m_\pi,y) \right) 
      -\frac{L|\vec{n}|f(m_\pi,y)}{3} \,K_1\left(L|\vec{n}| f(m_\pi,y) 
                                                        \right)\right] 
\nonumber \\
 & &{}-\frac{c_A^2g_A^0}{\pi^2F_\pi^2}\int_0^\infty dy\, y
{\sum_{\vec{n}}}' \left[K_0\left(L|\vec{n}| f(m_\pi,y) \right) 
       -\frac{L|\vec{n}|f(m_\pi,y)}{3} \,K_1\left(L|\vec{n}| f(m_\pi,y) 
                                                       \right)\right] 
\nonumber \\
 & &{}+\frac{8c_A^2g_A^0}{27\pi^2F_\pi^2}\int_0^\infty dy
{\sum_{\vec{n}}}' \frac{f(m_\pi,y)^2}{\Delta_0}  
  \left[K_0\left(L|\vec{n}| f(m_\pi,y) \right) 
  -\frac{K_1\left(L|\vec{n}| f(m_\pi,y) \right)}{L|\vec{n}|f(m_\pi,y)}\right] 
\nonumber \\
 & &{}-\frac{4c_A^2g_A^0}{27\pi F_\pi^2}\,\frac{m_\pi^3}{\Delta_0}
{\sum_{\vec{n}}}' \frac{1}{L|\vec{n}| m_\pi}\,e^{-L|\vec{n}| m_\pi}
                                                    +{\mathcal O}(\epsilon^4)  
\label{Deltaga}
\end{eqnarray}
with
\begin{equation}
f(m_\pi,y)  = \sqrt{m_\pi^2+y^2+2y\Delta_0} \,.
\end{equation}
The \raisebox{0.9mm}{${\sum}'$} indicates the omission 
of the $|\vec{n}|=0$ contribution in 
the sum over all vectors $\vec{n}$ with integer components. 
The contributions of the individual Feynman diagrams shown
in Fig.~\ref{diagrams} to $\Delta g_A$ are given in Appendix A.
The axial coupling of the nucleon in a finite volume depends on the very 
same parameters as the infinite volume result. 

Note that the recent analysis of Ref.~\cite{MIT} is based on the 
calculation by Beane and Savage~\cite{beane}, which does not employ
decoupling constraints in the chiral extrapolation function of
$g_A(\infty)$. Presumably this results in a fit function where the 
coupling $g_A^0$ does not agree with the value of $g_A$ at vanishing pion
mass. This in turn could lead to different predictions for the finite 
volume corrections, since the analytical formula for
$\Delta g_A(L)$ should be independent of the use of decoupling constraints
but is expressed in terms of the same parameters as the infinite volume 
result. In Appendix A we attempt a comparison between our result for 
$\Delta g_A(L)$ of Eq.~(\ref{Deltaga}) and the calculation by 
Beane and Savage~\cite{beane}.

Equation~(\ref{gafinite}) now allows us to access a larger set of 
lattice data than the formula (\ref{gaSSEneu}), without introducing 
new parameters.

%%%%%%%%%%%%%%%%%%%%%%%%%%%%%%%%%%%%%%%%%%%%%
\section{Fit results}\label{fit}
%%%%%%%%%%%%%%%%%%%%%%%%%%%%%%%%%%%%%%%%%%%%%%%%

As in previous applications of SSE results to the chiral extrapolation 
of baryon properties (see, e.g., Ref.~\cite{HPW,HW,ff}) we 
do not have enough data points at sufficiently small
masses to fit all parameters. 
So we must fix some of the parameters at reasonable
values. What is known from phenomenology? The analysis of (inelastic) 
$\pi N$ scattering, in particular the process 
$\pi N \to \pi \pi N$~\cite{Nadia}, 
suggests that choosing the physical pion mass as the scale $\lambda$ 
one has~\cite{HPW} 
\begin{equation} \label{bphen}
B_9^r(\lambda = m_\pi^{\mathrm {phys}}) = (-1.4 \pm 1.2) \,\mbox{GeV}^{-2}
\quad , \quad B_{20}^r(\lambda = m_\pi^{\mathrm {phys}}) \equiv 0 \,.
\end{equation}
The coupling $B_{20}^r$ cannot be observed independently of $B_9^r$, 
as it accompanies a structure in the chiral Lagrangian which
is proportional to the equation of motion (see e.g.\ Ref.~\cite{Ecker}). 
Hence the separation between the two couplings given by Eq.~(\ref{bphen}) 
can be utilized without imposing any model assumptions.

Therefore we set $\lambda = 0.14 \, \mbox{GeV}$ in the following and 
take $B_{20}^r$ to be zero in order to utilize this valuable information 
from scattering experiments. Furthermore, from analyses of 
$\pi N$ scattering and $\pi$-photoproduction in the $\Delta$(1232) 
resonance region one knows~\cite{PDG} the real part of the 
$N \Delta$ mass splitting to be $0.271 \, \mbox{GeV}$ at the physical point. 
We note that a recent analysis~\cite{BHM} of the quark-mass dependence 
of the (real part of the) mass of $\Delta$(1232) found a slightly 
higher value for the splitting between the mass of the nucleon and its 
first excited state in the chiral limit, $\Delta_0\approx 330$ MeV, 
albeit with a large uncertainty. We will study the implications of this 
finding in a future communication~\cite{future}. The imaginary 
part of the complex mass of $\Delta$(1232) is also
known~\cite{PDG}: In an  $O(\epsilon^3)$ analysis within SSE it corresponds
to a strong decay into $\pi N$ intermediate states 
with the coupling value $c_A = 1.5$~\cite{GH}. In the previous analysis 
of the quark-mass dependence of $g_A$ of Ref.~\cite{HPW} a smaller value 
for $c_A$ was utilized. 
Ultimately such an issue can only be decided by 
performing simultaneous global fits to several observables sensitive 
to $\Delta$(1232) contributions within the same ChEFT formalism. 

At the physical pion mass we also know the very precise value for the 
axial coupling  $g_A^{\mathrm {exp}} = 1.2695$ of the nucleon from 
neutron beta decay analyses~\cite{PDG}, which in the 
${\cal O}(\epsilon^3)$ SSE analysis of Ref.~\cite{HPW} 
together with a set of quenched lattice data
led to the estimate $g_A^0 \approx 1.2$. Little is known about
$g_1$. In the SU(6) quark model one would expect 
$g_1 = \frac{9}{5} g_A^0 \approx \frac{9}{5} 1.2 = 2.16 $. For the 
pion decay constant ChPT analyses suggest a chiral limit value of 
$F_\pi \approx 86.2 \, \mbox{MeV}$~\cite{Colangelo}.
In the following we fix $\Delta_0 = 0.271 \, \mbox{GeV}$, $c_A = 1.5$,
$F_\pi = 86.2 \, \mbox{MeV}$ and leave $g_A^0$, $g_1$, 
$B_9^r \equiv B_9^r(\lambda = 0.14 \, \mbox{GeV})$ as fit parameters.

To be on the safe side as far as the applicability of ChEFT is concerned
we start with a fit that takes into account only the six
data points at our lowest quark masses (data points 3, 6, 12, 13, 
14 and 17 in Tables~\ref{tab:param} and \ref{tab:gabare}). 
Indeed, we cannot expect ChEFT to work
at pion masses well above $500-600 \, \mbox{MeV}$,
because the pion-loop integrals start to become more and more 
sensitive to scales beyond $\Lambda_\chi \sim 4\pi F_\pi$ for such large 
masses (see the discussion in Ref.~\cite{cutoff} for details).
Fortunately, we can include one ($\beta$, $\kappa$) pair where 
we have simulations for three different volumes. 

Thus we deal with a three-parameter fit of six data points, which we 
shall call Fit A. In the fit function we have, of course, to 
include the finite volume correction, i.e.\ we use 
$g_A(L) = g_A(\infty) + \Delta g_A(L)$ with $g_A(\infty)$
given in Eq.~(\ref{gaSSEneu}) and $\Delta g_A(L)$ taken from 
Eq.~(\ref{Deltaga}). The resulting values for the fit parameters 
are given in Table~\ref{tab:fitresults}. 

\begin{table*}
\caption{Fit results.} 
\label{tab:fitresults}
\begin{ruledtabular}
\begin{tabular}{clllc}
fit & \multicolumn{1}{c}{$g_A^0$} & \multicolumn{1}{c}{$B_9^r$} &
 \multicolumn{1}{c}{$g_1$} & $\chi^2$/dof \\ \hline
A &  1.15(12) & $- 0.71(18) \, \mbox{GeV}^{-2}$ & 2.6(8) & 1.41 \\
B &  1.26(8)  & $- 0.62(3)  \, \mbox{GeV}^{-2}$ & 3.3(7) & 1.54 \\
\end{tabular}
\end{ruledtabular}
\end{table*}

\begin{figure}[htb]
\begin{center}
\epsfig{file=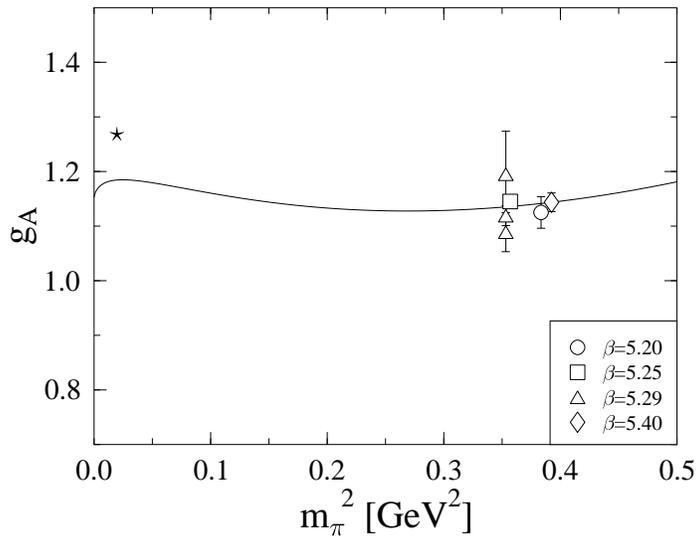,width=10cm} 
\end{center}
\vspace*{-0.3cm}
\caption{Fit A of the $g_A$ data.
The finite size correction 
has been subtracted from the simulation results. Only data points
included in the fit are shown. The star represents the physical point.}
\label{fig:gafit}
\end{figure}

In Fig.~\ref{fig:gafit} we plot the data
with the finite size correction $\Delta g_A (L)$ subtracted 
together with the fit curve.
If the fit would describe the volume dependence of the data perfectly,
the data points from simulations differing only in the volume 
(simulations 12, 13 and 14) would fall on top of each other 
in the plot. As can be seen in Fig.~\ref{fig:gafit} this is indeed 
rather well satisfied within the error bars. Furthermore,
subtracting $\Delta g_A (L)$ has moved the ``large volume'' results
at our smallest quark masses from all four $\beta$ values even closer 
together, corroborating our previous impression that lattice artefacts
are small.

\begin{figure}[htb]
\begin{center}
\epsfig{file=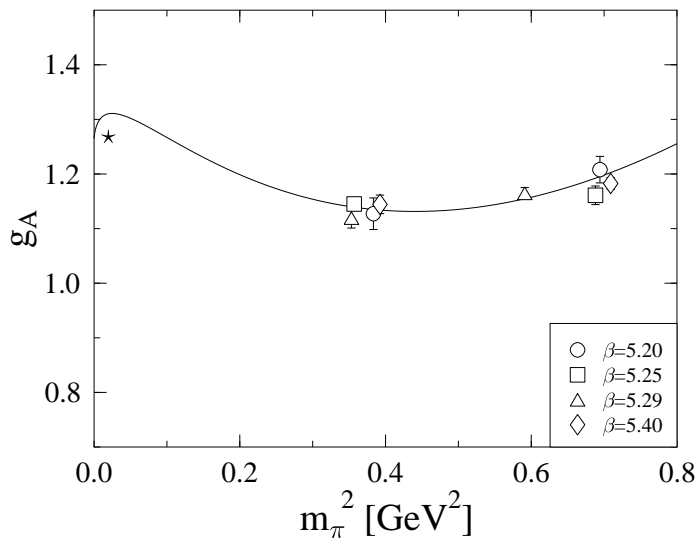,width=10cm}
\end{center}
\vspace*{-0.3cm}
\caption{Fit of the $g_A$ data over an extended range of pion masses (Fit B).
The finite size correction has been subtracted from the data.
In contrast to Fig.~\ref{fig:gafit}, only the resulting infinite volume 
numbers are shown for the two masses at $\beta = 5.29$. 
The three data points around $m_\pi^2 \approx 0.7 \, \mbox{GeV}^2$ 
have not been included in the fit.} 
\label{fig:gafitext}
\end{figure}

Remarkably enough, our fits do not break down when data at somewhat 
higher pion masses are included. Indeed, if we take 
into account also the data from simulations 9, 10 and 11, where 
$m_\pi \approx 0.77 \, \mbox{GeV}$, we obtain the results labeled as Fit B
in Table~\ref{tab:fitresults}. They are well compatible with the numbers
from Fit A. In Fig.~\ref{fig:gafitext} the fit curve is
confronted with the data.

\begin{figure}
\begin{center}
\epsfig{file=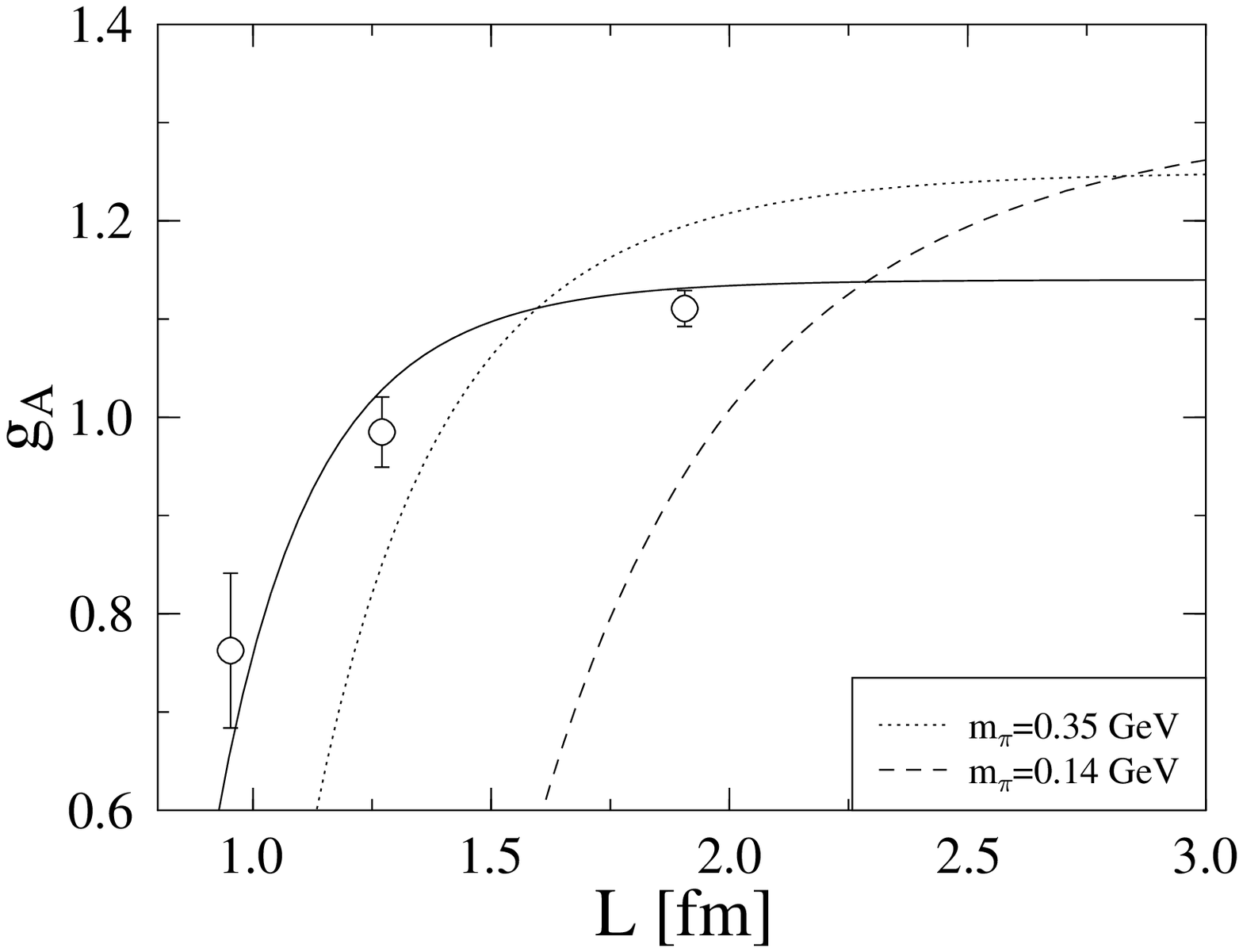,width=10cm} \\
\epsfig{file=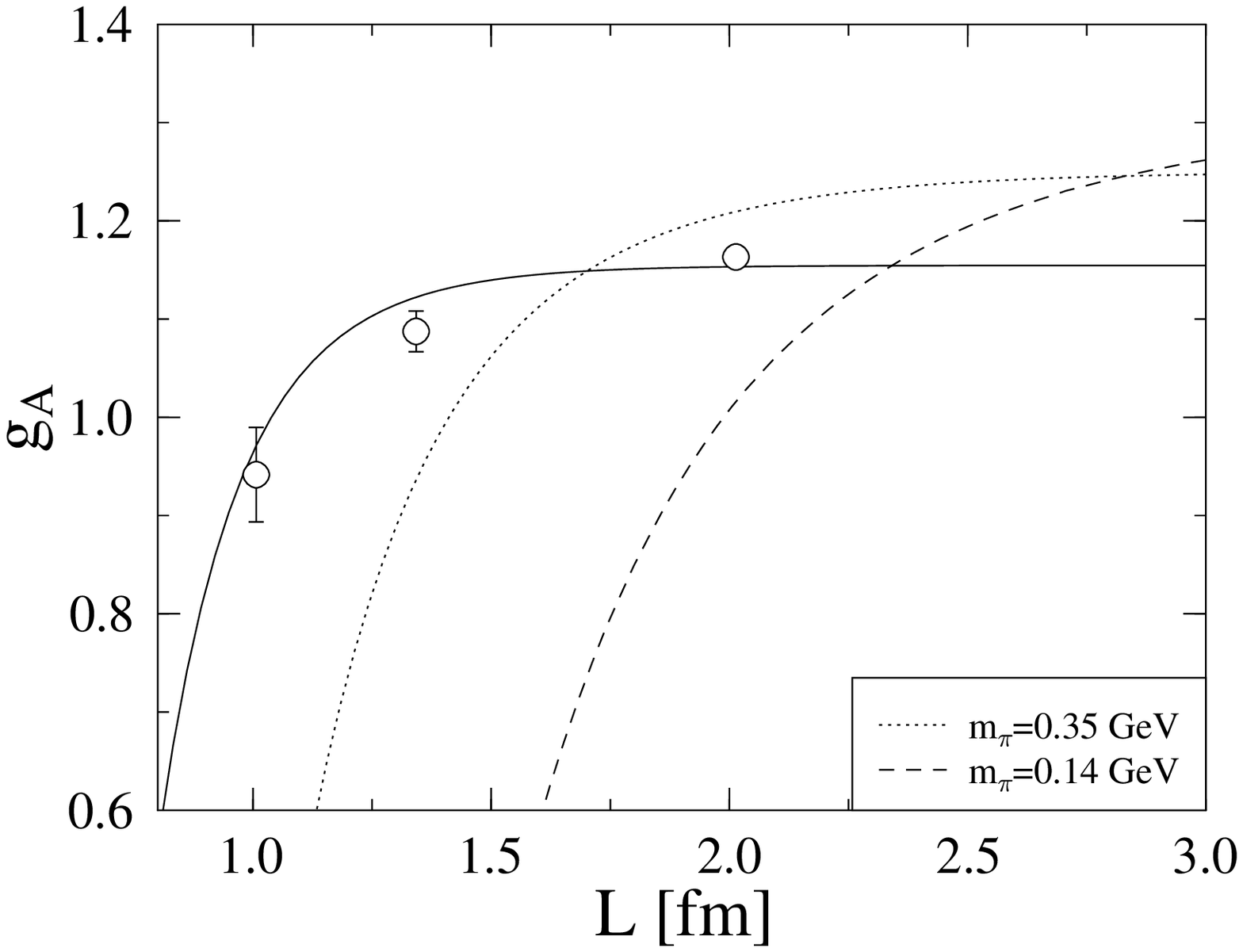,width=10cm}
\end{center}
\caption{Results for $g_A$ at $\beta = 5.29$, $\kappa = 0.1359$ 
(corresponding to a pion mass of about $0.6 \, \mbox{GeV}$, upper plot) 
and at $\beta = 5.29$, $\kappa = 0.1355$ 
(corresponding to a pion mass of about $0.77 \, \mbox{GeV}$, lower plot) 
plotted versus $L$. The full curves have been computed for the corresponding 
pion masses using the parameters from Fit B. 
The dotted (dashed) curves show the volume dependence 
expected from these parameters for $m_\pi = 0.35 \, \mbox{GeV}$
($m_\pi = 0.14 \, \mbox{GeV}$).} 
\label{fig:gavsl}
\end{figure}

Using the fit parameters from Fit B, we plot in Fig.~\ref{fig:gavsl} the
dependence of $g_A$ on the box size $L$ for $m_\pi = 0.594 \, \mbox{GeV}$ 
together with our three data points at $\beta = 5.29$, $\kappa = 0.1359$
and for $m_\pi = 0.769 \, \mbox{GeV}$ along with the three data points 
at $\beta = 5.29$, $\kappa = 0.1355$.
In addition we show the behavior to be expected at 
$m_\pi = 0.35 \, \mbox{GeV}$ and at the physical pion mass.
These plots show clearly how well Fit B reproduces the volume 
dependence of our data. 

\begin{figure}
\begin{center}
\epsfig{file=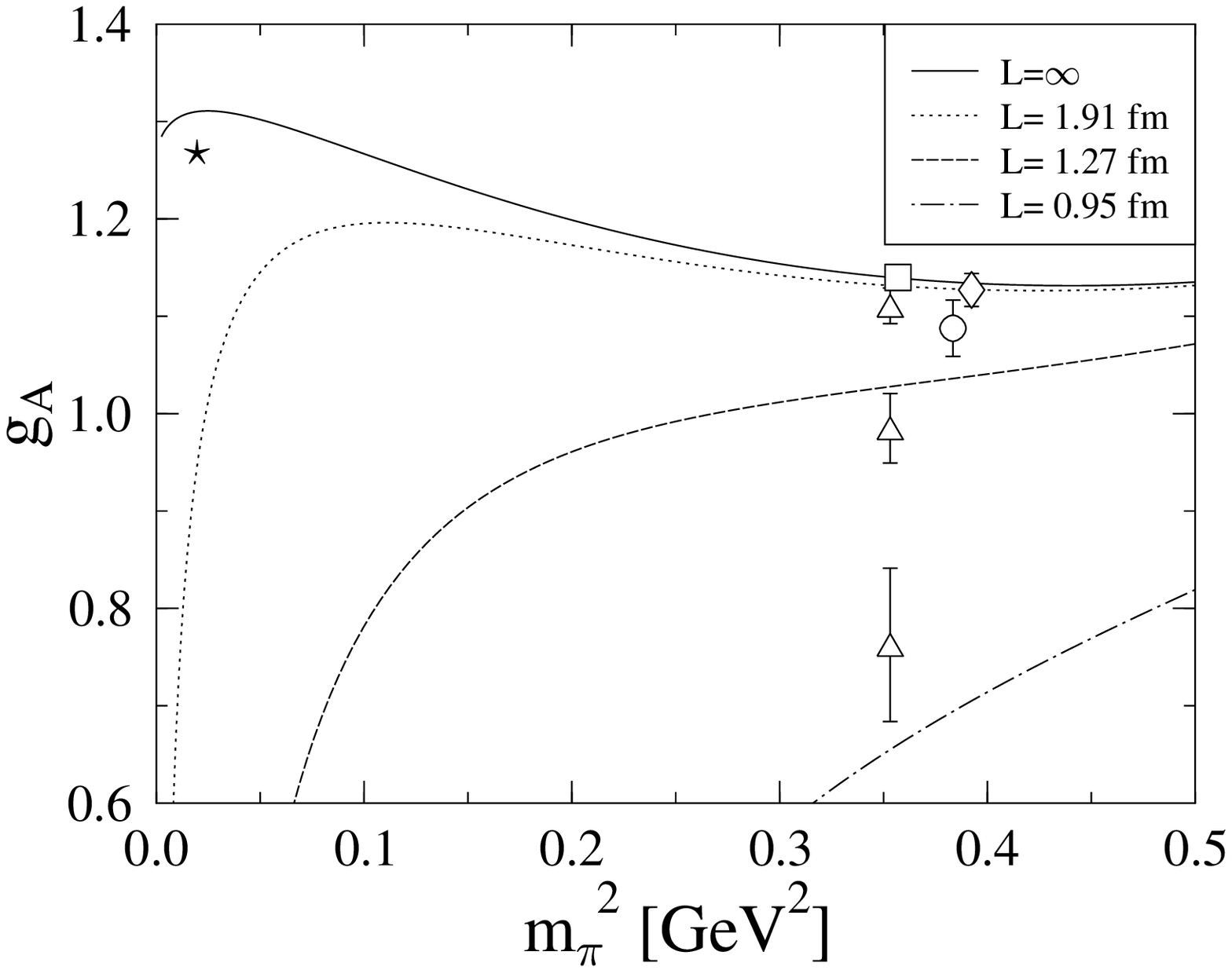,width=10cm}
\end{center}
\caption{Results for $g_A$ in the simulated volumes together with
fit curves from Fit B.} 
\label{fig:gafinvol}
\end{figure}

In Fig.~\ref{fig:gafinvol} we display the volume dependence of the 
data and the fit in yet another way. There we plot our results as 
they were obtained in the respective volumes versus $m_\pi^2$. For the 
curves in this plot we take the parameters from Fit B. The finite $L$
values correspond to the volumes used in the simulations 12, 13, 14.

In order to study the influence of the smaller volumes 
(simulations 9, 10 and 12, 13) we have repeated Fit B
without these small-volume data, i.e.\ using only the results 
from simulations 3, 6, 11, 14 and 17. This yields $g_A^0 = 1.21 \pm 0.19$,
$B_9^r = (-0.62 \pm 0.06) \mbox{GeV}^{-2}$ and $g_1 = 2.8 \pm 1.7$. 
Of course, the errors have increased, but the fitted parameters are 
fully compatible with the outcome of Fit B. 

In order to estimate the uncertainty caused by the
ambiguities inherent in the scale setting procedure (see Sect.~\ref{simu})
we have not only employed $r_0=0.467 \, \mbox{fm}$ in the fits but also 
$r_0 = 0.5 \, \mbox{fm}$. Moreover, we have considered the two
possibilities of working with $r_0/a$ evaluated at the respective quark
mass and using the chirally extrapolated value of $r_0/a$. For 
$g_A^0$ we find numbers between 1.12 and 1.21 in Fit A, while
they vary between 1.19 and 1.33 in Fit B. The results for $B_9^r$ 
(in $\mbox{GeV}^{-2}$) lie between $-0.60$ and $-0.83$ in Fit A 
and between $-0.54$ and $-0.72$ in Fit B. The values for 
$g_1$ range between 2.1 and 3.2 in Fit A and between 2.5 
and 4.1 in Fit B. We shall take half of these spreads 
as our estimate of the systematic error due to setting the scale.

Let us now discuss the values that our fits have yielded for the parameters. 
For the axial coupling in the chiral limit $g_A^0$ we find a result
below the value at the physical point.
This is in agreement with the analyis of Ref.~\cite{HPW} 
based on the quenched data of QCDSF~\cite{gaquenched}. 
The small negative value for the 
coupling $B_9^r$ is also entirely within expectations from the 
analyses of inelastic $\pi N$-scattering \cite{HPW,Nadia} 
(see Eq.~\ref{bphen}). Finally, we note that the value for the 
axial $\Delta\Delta$ coupling $g_1$ in this new analysis is much 
closer to the SU(6) quark model result than the number 
found in Ref.~\cite{HPW}, reassuring us that the contributions 
of $\Delta$(1232) to the axial coupling of the nucleon are under control. 
The main cause for this more reasonable value of $g_1$ is the 
larger value for the coupling $c_A$ utilized in the present study~\cite{tim}. 

We note that the curves for $g_A$ in a finite volume presented 
previously in Ref.~\cite{Lattice04}---which were consistently 
above our infinite-volume curve---can be obtained from the 
present analyis by setting all $\Delta(1232)$ contributions in 
$\Delta g_A(L)$ identically to zero. The complete finite volume effects 
found here within ${\cal O}(\epsilon^3)$ SSE, in contrast, are 
negative at large quark masses, consistent with the findings 
from the simulation. The resulting finite-volume corrections therefore 
arise from a destructive interference between $N\pi$ and $\Delta\pi$  
loop effects. This point will be discussed in more detail in a future 
publication~\cite{future}. Such a strong 
cancellation between Goldstone Boson loops around octet and decuplet
baryon intermediate states has also been observed in Ref.~\cite{JM} 
in an SU(3) HBChPT calculation of the axial current of the nucleon 
in infinite volume, albeit under the additional assumption of 
degenerate octet-decuplet multiplets.

%%%%%%%%%%%%%%%%%%%%%%%%%%%%%%%%%%%%
\section{Summary}
%%%%%%%%%%%%%%%%%%%%%%%%%%%%%%%%%%%%%%%

We have evaluated the axial coupling constant $g_A$ of the nucleon in lattice 
QCD on gauge field configurations generated with two degenerate 
flavors of dynamical non-perturbatively improved Wilson fermions. 
The necessary renormalization of the axial vector current has 
been performed non-perturbatively within the RI$^\prime$-MOM 
scheme, except for the mass-dependent factor: $b_A$ has been 
calculated in tadpole improved boosted perturbation theory.

For two ($\beta$, $\kappa$) combinations we have performed simulations
on three different volumes allowing us to study finite size effects.
For the chiral extrapolation as well as for the description of the 
volume dependence we have made use of ChEFT.
For this purpose we have rewritten the expression for the quark-mass 
dependence of $g_A$ derived within the framework of the SSE
in Ref.~\cite{HPW} in a form which is particularly suitable 
for our application, and we have extended the calculation to cover also
the volume dependence. With this formula at hand, we could perform
a fit to our data obtained on lattices of different spatial extent.
Note, however, that we had to fix some parameters at 
phenomenologically reasonable values. Then finite volume effects 
are reproduced surprisingly well down to box lengths of about 
$1 \, \mbox{fm}$. Nevertheless, simulations at smaller quark 
masses will be necessary to confirm our findings.

In Ref.~\cite{jaffe} it has been argued that, in the chiral limit, 1/3 of
the axial charge of the nucleon is to be found at infinite distance from
the nucleon, due to a delocalization effect. Furthermore, it was suggested
that this phenomenon could lead to large finite volume effects. While
there is agreement on the delocalization phenomenon, it has been disputed
that this effect will cause calculations of $g_A$ in a finite system to
have particularly large finite volume effects~\cite{cohen}. Notwithstanding, 
our fits suggest, for example, that at the physical pion mass the 
axial charge is reduced by a factor $\approx 2$ if the nucleon is 
confined to a periodic box of extent $L \approx 1.5 \, \mbox{fm}$. 
This result is in good agreement with a model calculation~\cite{TALY}, 
in which the effect is attributed to periodic boundary conditions 
rather than delocalization of the axial charge.

Given that the SSE is based upon the long distance pion dynamics 
around nucleon and $\Delta$(1232) matter states,
we conclude that contributions from $\Delta$(1232) are crucial 
in understanding the quark-mass (and volume) dependence of $g_A$. 
This confirms the findings of Ref.~\cite{HPW} and 
the old expectations (at the physical point) based on the 
observation that the integral in
the Adler-Weisberger sum rule, which represents $g_A^2 - 1$, is 
dominated by the $\Delta$(1232) resonance.
The values for the couplings involving
$\Delta$(1232) dynamics employed in our chiral extrapolation 
curve are consistent both with scattering phenomenology and
with the expectations of the SU(6) quark model.
In particular the value for the axial $\Delta\Delta$ SSE coupling 
$g_1\approx 3.0$ in this new analysis is much more consistent with 
expectations from phenomenology than the value obtained in Ref.~\cite{HPW}.

Inclusion of simulation data at several values of the box length $L$ 
is crucial to increase the number of data points at sufficiently small
pion masses. The two-dimensional surface $(m_\pi, L)$ allows for a much 
better determination of the individual values of the effective 
couplings. With the new dynamical simulation data presented here and our 
${\cal O}(\epsilon^3)$ SSE analysis we can extract the chiral limit
value $g_A^0$ from the extrapolation without including any 
constraints on the physical point. 
As our final result we quote $g_A^0$ from Fit B, which takes into
account pion masses below $800 \, \mbox{MeV}$. We find 
\begin{equation}
g_A^0 = 1.26(8)(7) \,. 
\end{equation}
The first error is statistical, while the second error is an estimate 
of the systematic uncertainty caused by the ambiguities in setting 
the scale. Repeating the fit with the value of $g_A$ at the 
physical pion mass as fit parameter instead of $g_A^0$ we obtain
\begin{equation}
g_A(m_\pi^{\mathrm {phys}}) = 1.31(9)(7) \,.
\end{equation}

Ultimately, one would like to determine all parameters of ChEFT solely
from a fit to lattice data. Such an enterprise would require a joint
fit of results from simulations with dynamical quarks for many static nucleon 
and $\Delta$(1232) observables, which presently is out of reach due 
to the paucity of lattice data for pion masses below $600 \, \mbox{MeV}$ 
in many of these observables.

\section*{Acknowledgements}

The numerical calculations have been performed on the Hitachi SR8000
at LRZ (Munich), on the Cray T3E at EPCC (Edinburgh)~\cite{allton},
and on the APEmille at NIC/DESY
(Zeuthen). This work is supported in part by the DFG (Forschergruppe
Gitter-Hadronen-Ph\"anomenologie and Emmy-Noether program) and by the EU
Integrated Infrastructure Initiative Hadron Physics under contract
number RII3-CT-2004-506078.

A.A. thanks the ``Berliner Programm zur F\"orderung der
Chancengleichheit f\"ur Frauen in Forschung und Lehre''
for financial support.
T.R.H. acknowledges the hospitality of the Forschergruppe 
``Gitter-Hadronen-Ph\"anomenologie'' at the University of Regensburg 
and at DESY Zeuthen. 

\newpage
%%%%%%%%%%%%%%%%%%%%%%%%%%%%%%%%%%%%%%%%%%%%%%%%%%%%%%%%%%%%%%%%%
\appendix
%%%%%%%%%%%%%%%%%%%%%%%%%%%%%%%%%%%%%%%%%%%%%%%
\section{Amplitudes in finite volume}
%%%%%%%%%%%%%%%%%%%%%%%%%%%%%%%%%%%%%%%%%

The amplitudes of the contributions of the eight Feynman diagrams 
of Fig.~\ref{diagrams} to $g_A$ in the infinite volume can be found in the
appendix of Ref.~\cite{BFHM}. Here we give the individual contributions 
of the eight diagrams to $\Delta g_A$ defined in Eq.~(\ref{gafinite}):

\begin{eqnarray}
\Delta Amp_1&=& \mathrm i(\eta^{\dagger}\tau^b\eta)\times\bar{u}(r_1)
 S\cdot\epsilon_Au(r_2) \nonumber\\
  &&\times \Bigg[\frac{9(g_A^0)^3}{4F_{\pi}^2} \left(\frac{m_{\pi}^2}{12\pi^2}
{\sum_{\vec{n}}}^{\prime}K_0(L|\vec{n}|m_{\pi})-
  \frac{m_{\pi}}{12\pi^2L}{\sum_{\vec{n}}}^{\prime}\frac{1}{|\vec{n}|}
           K_1(L|\vec{n}|m_{\pi})\right)\nonumber\\
 &&-\frac{c_A^2g_A^0}{\pi^2F_{\pi}^2}\int_{0}^{\infty} \! \mathrm dy \, y
  {\sum_{\vec{n}}}^{\prime}\Bigg(K_0(L|\vec{n}|
              \sqrt{m_{\pi}^2+y^2+2y\Delta_0})\nonumber\\
 &&-\frac{1}{3}L|\vec{n}|\sqrt{m_{\pi}^2+y^2+2y\Delta_0}
      K_1(L|\vec{n}|\sqrt{m_{\pi}^2+y^2+2y\Delta_0})\Bigg)\Bigg] \,,
\label{zfac1}\\
 \Delta Amp_2&=&\Delta Amp_3=0 \,, \\
 \Delta Amp_4&=&- \mathrm i (\eta^{\dagger}\tau^b\eta)\times\bar{u}(r_1)
  S\cdot\epsilon_A u(r_2)\frac{(g_A^0)^3}{4F_{\pi}^2} \nonumber\\
 && \times \left(\frac{m_{\pi}^2}{12\pi^2}
    {\sum_{\vec{n}}}^{\prime}K_0(L|\vec{n}|m_{\pi})
    -\frac{m_{\pi}}{12\pi^2L}{\sum_{\vec{n}}}^{\prime}\frac{1}{|\vec{n}|}
                               K_1(L|\vec{n}|m_{\pi})\right) \,, \\
\Delta Amp_5&=&- \mathrm i(\eta^{\dagger}\tau^b\eta)\times
  \bar{u}(r_1) S\cdot\epsilon_Au(r_2)\frac{g_A^0m_{\pi}}{4\pi^2F_{\pi}^2L}
    {\sum_{\vec{n}}}^{\prime}\frac{1}{|\vec{n}|}K_1(L|\vec{n}|m_{\pi}) \,, \\
\Delta Amp_6&=&\Delta Amp_7\nonumber\\
    &=& \mathrm i(\eta^{\dagger}\tau^b\eta)\times
                          \bar{u}(r_1)S\cdot\epsilon_Au(r_2)
       \frac{4g_A^0c_A^2}{27\pi^2\Delta_0 F_{\pi}^2}\nonumber\\
    &&\times \Bigg[\int_{0}^{\infty} \! \mathrm dy {\sum_{\vec{n}}}^{\prime}
       \Bigg(\sqrt{m_{\pi}^2+y^2+2y\Delta_0}\nonumber\\
    &&\times \bigg(-\frac{1}{L|\vec{n}|}K_1(L|\vec{n}|
                          \sqrt{m_{\pi}^2+y^2+2y\Delta_0})
     +\sqrt{m_{\pi}^2+y^2+2y\Delta_0} \nonumber\\
    &&\times K_0(L|\vec{n}|\sqrt{m_{\pi}^2+y^2+2y\Delta_0})\bigg)\Bigg)
       -\frac{m_{\pi}^2\pi}{2}{\sum_{\vec{n}}}^{\prime}
                   \frac{1}{L|\vec{n}|}e^{-L|\vec{n}|m_{\pi}}\Bigg] \,, \\
\Delta Amp_8&=& \mathrm i (\eta^{\dagger}\tau^b\eta)\times\bar{u}(r_1)
  S\cdot\epsilon_Au(r_2)\frac{25}{81}
                    \frac{c_A^2g_1}{\pi^2F_{\pi}^2}\nonumber\\
    &&\times \int_{0}^{\infty} \! \mathrm dy \, y{\sum_{\vec{n}}}^{\prime}
          \Bigg(K_0(L|\vec{n}|\sqrt{m_{\pi}^2+y^2+2y\Delta_0})\nonumber\\
    &&-\frac{1}{3}L|\vec{n}|\sqrt{m_{\pi}^2+y^2+2y\Delta_0}
                K_1(L|\vec{n}|\sqrt{m_{\pi}^2+y^2+2y\Delta_0})\Bigg) \,.
\label{zfac2}
\end{eqnarray}
Additional details are given in Ref.~\cite{tim}.
Note that the finite-volume shift in diagram 1 arises from the 
$Z$ factor of the nucleon. In Eqs.~(\ref{zfac1}) - (\ref{zfac2})
$S_\mu$ denotes the Pauli-Lubanski vector of ChEFT, whereas 
$\epsilon_A^\mu$ corresponds to an external axial-vector background 
source interacting with the hadronic system; $u(r_1),\,\bar{u}(r_2)$ 
are the non-relativistic spinors of the incoming/outgoing nucleon with 
4-momenta $r_1^\mu,\,r_2^\mu$ respectively. The bilinear combination 
$\eta^\dagger \tau^b \eta$ with $b=1,2,3$ denoting the 
isovector index of the background source encodes the isospin dependence 
of the current. The sums extend over all vectors $\vec{n}$ with integer
components excluding $\vec{n} = \vec{0}$. For further details on the 
notation we refer to Ref.~\cite{BFHM}. The couplings and parameters 
occurring in these relations are defined in the main text.

In Table~\ref{xxx} we have attempted to relate our couplings 
$g_A^0$, $c_A$, $g_1$, $\Delta_0$ and $F_\pi$ to the ones used by 
Beane and Savage in Ref.~\cite{beane}. However, due to the 
decoupling constraints of the SSE scheme (see the discussion in 
Sec.~\ref{sec:cheft}) the correspondence between the two sets of couplings 
shown in Table~\ref{xxx} is strictly true only at leading order 
in the chiral expansion.

While we are also employing a different representation of the 
finite-volume shifts in terms of the Bessel functions, numerically 
we can reproduce the results shown in Ref.~\cite{beane} by utilizing 
the two parameter sets discussed there and shown in 
Table~\ref{yyy}. However, the sign of the finite-volume 
shift observed in the Monte-Carlo simulation of $g_A$ reported in 
this work does not agree with either of these two coupling scenarios, 
which would both lead to an increase of $g_A$ when the volume is 
decreased for pion masses above 140 MeV. Finally, we emphasize again 
that our set of couplings of the effective theory in the SSE scheme 
has been obtained for the complete result $g_A(L)$ of 
Eq.~(\ref{gafinite}), and not just for the 
finite-volume shift $\Delta g_A(L)$ as discussed in Ref.~\cite{beane}. 
At ${\cal O}(\epsilon^3)$, differences between the SSE scheme used 
in this work and the approach of Ref.~\cite{beane}---aside from 
numerical differences in the couplings---only manifest themselves in 
the $g_A(\infty)$-part of $g_A(L)$.

\begin{table*}
\caption{The correspondence (at leading order) between the SSE 
couplings used in this work and the parameters used in 
Ref.~\cite{beane}.} 
\label{xxx}
\begin{ruledtabular}
\begin{tabular}{cc}
SSE-couplings & Couplings of Ref.~\cite{beane} \\ \hline
    $g_A^0$ & $g_A^{BS}$\\
    $g_1$ & $-g_{\Delta\Delta}^{BS}$\\
    $c_A$ & $-\frac{g_{\Delta N}^{BS}}{\sqrt{2}}$\\
    $\sqrt{2}F_{\pi}$ & $ f_{BS}$\\
    $\Delta_0$ & $\Delta_{BS}$ \\
\end{tabular}
\end{ruledtabular}
\end{table*}

\begin{table*}
\caption{The two coupling scenarios discussed in Ref.~\cite{beane}.}
\label{yyy}
\begin{ruledtabular}
\begin{tabular}{ccc}
Parameter & (a)& (b)\\ \hline 
  $g_A^{BS}$ & 1.67 &1.33\\
  $\Delta_{BS}$ & 0.293 GeV&0.293 GeV\\ 
  $g_{\Delta N}^{BS}$ & -2&-1.41\\ 
  $g_{\Delta\Delta}^{BS}$ & -3&-3\\ 
  $f_{BS}$ & 0.132 GeV &0.132 GeV
\end{tabular}
\end{ruledtabular}
\end{table*}

%%%%%%%%%%%%%%%%%%%%%%%%%%%%%%%%%%%%%%%%%%%%%%%%%%%%%%%
%%%%%%%%%%%%%%%%%%%%%%%%%%%%%%%%%%%%%%%%%%%%%%%%%%%%%%%%

\end{document}